\documentclass[
aip,
a4paper,
jmp,
 amsmath,amssymb,
 reprint,%
author-numerical,%
]{revtex4-2}

\usepackage{graphicx}
\usepackage{dcolumn}
\usepackage{bm}
\usepackage{natbib}
\setcitestyle{numbers,square}
\usepackage[hidelinks]{hyperref}
\hypersetup{colorlinks=true,allcolors=[rgb]{0,0,1}}
\usepackage[center]{titlesec}
\titleformat*{\section}{\centering\normalsize\bfseries}
\titleformat*{\subsection}{\centering\normalsize\bfseries}
\usepackage[a4paper, portrait,margin=23mm]{geometry}

\begin{document}
\title{On the stability of large-amplitude gravity-capillary surface waves}
\author{Josh Shelton}
 \email{josh.shelton@st-andrews.ac.uk}
\affiliation{%
School of Mathematics and Statistics, University of St Andrews, St Andrews, KY16 9SS, UK
}
\author{Adam Rook}
\affiliation{
 Department of Mathematical Sciences, University of Bath, Bath, BA2 7AY, UK
}
\fontsize{12}{14}\selectfont
\begin{abstract}
We consider the stability of periodic gravity-capillary waves of finite amplitude for small values of the surface tension.
Linear stability with respect to both superharmonic and subharmonic perturbations is calculated for each solution, and our methodology obtains the full eigenvalue spectrum consisting of growth rates and temporal frequencies. 
For small surface tension, the gravity-capillary wave solution space consists of a countably-infinite number of solution branches that coalesce in the small-surface-tension limit, which forms one of the main complications of our study.
When the energy is fixed as an amplitude constraint, we find that the superharmonic instability associated with near-limiting gravity waves emerges at smaller amplitudes in the presence of surface tension.
Further, the modulational (long-wave) instability is seen to be stabilised for finite-amplitude solutions in the presence of surface tension.
This occurs at surface tension values well below that previously obtained via weakly-nonlinear theory, and the stabilisation is nonmonotonic as very small fluctuations in the surface tension of solutions produce large changes in their stability properties.

\end{abstract}
\maketitle
\section{Introduction}
\label{sec:intro}
\fontsize{12}{14}\selectfont

\noindent Consider the behaviour of a travelling wave on the surface of a fluid subject to the effect of gravity.
Travelling gravity waves of large amplitude have high curvature at the wave crest, and this curvature is known to tend to infinity as the limiting Stokes wave is approached.
However, the effect of surface tension on the dynamics of the wave is proportional to the surface curvature.
Therefore in a regime with even a very small amount of surface tension this effect becomes increasingly relevant at high amplitudes.

The instability of periodic surface waves is known to be one of the primary mechanisms by which large-amplitude waves are generated in the ocean \cite{benjamin1967disintegration}.
While the stability of large-amplitude gravity waves has received extensive study, very little is known of the analogous case that includes surface tension except for results in the small-amplitude limit.
In this work we perform a systematic investigation of the stability of large-amplitude gravity-capillary waves, and demonstrate the stabilising effect that surface tension can have on long-wave temporal perturbations.

Multiple different instability mechanisms exist for periodic gravity waves.
Small-amplitude gravity waves are known to be modulationally unstable to long-wave perturbations \cite{benney1967propagation}, and this is often referred to as the Benjamin-Feir instability \cite{benjamin1967disintegration}.
The subharmonic instability of finite-amplitude gravity waves, corresponding to perturbations of incommensurate wavelength to the travelling wave, was first shown by Longuet-Higgins \cite{10.1098/rspa.1978.0081}.
The limiting Stokes wave has an amplitude of $\mathcal{A}\approx 0.0705$, defined by half of the ratio of vertical extent to wavelength, and modulational stability occurs for $\mathcal{A}>0.0545$ \cite{deconinck2011instability}.
Further, gravity waves are known to become unstable to superharmonic perturbations for $\mathcal{A}>0.069$ \cite{tanaka1983stability,deconinck2024self}, corresponding to a turning point in the graph of energy against amplitude, and this forms the dominant growth rate for near-limiting profiles.

One of the main complications involved in our study is the complexity of the underlying solution space for travelling gravity-capillary waves.
In the absence of surface tension, there is a one-parameter family of periodic gravity waves.
However, the extension of this family of gravity-wave solutions to non-zero surface tension results in the formation of a complicated structure of solution branches due to the singular nature of the small-surface-tension limit.
When the energy is fixed as an amplitude constraint there exists a single gravity wave solution, from which a countably infinite number of solution branches sequentially emerge as the surface tension increases \cite{shelton2021structure}.
The result is a snaking bifurcation structure that coalesces in the limit of zero surface tension, and consists of solutions that contain oscillatory capillary modes.
Analogous gravity-capillary bifurcation structures have since been obtained for time-periodic standing waves \cite{shelton2023structure} and for waves subject to viscosity and forcing \cite{shelton2025time}.

Stability results for gravity-capillary waves have been obtained in the limit of small amplitude by several authors.
By studying weakly-nonlinear wave equations for both solitary and periodic waves, transverse instabilities were derived by Ablowitz \& Segur \cite{ablowitz1979evolution}.
More recently, there has been a focus on proof-based methods that have demonstrated the stabilisation of long-wave modes due to surface tension \cite{hur2023unstable,sun2025spectral}.
For finite-amplitude travelling waves, stability with respect to three-dimensional perturbations has been considered \cite{zhang1987three}, and it was found that instabilities were most prominent near solutions associated with resonance conditions.
The most related work to our present study was performed by Deconinck \& Trichtchenko \cite{deconinck2014stability}, who calculated the stability of gravity-capillary waves at fixed surface tension for increasing amplitude and select values of the fluid depth.
They discovered that the growth rate of instabilities associated with gravity waves become larger when surface tension is included.
However, their study contains several issues related to the steady solution space of gravity-capillary waves.
Firstly, restrictions on the number of Fourier modes considered meant that bifurcations between adjacent branches of steady solutions were not resolved.
These bifurcations emerge at fixed surface tension as the amplitude increases, which has been derived analytically (c.f. figure~12 from Shelton \& Trinh \cite{shelton2022exponential}), and occur at very high Fourier coefficients.
Secondly, no oscillatory capillary modes were resolved in their computations.
These are a critical feature of all solution profiles for small but non-zero surface tension, and were omitted due to the combination of a very small value of surface tension combined with fixed double-precision resolution of the steady solutions.
Such oscillatory modes would have appeared in the computations of \cite{deconinck2014stability} at higher amplitudes if the associated solution branches were resolved.

We consider the stability of periodic travelling waves in the gravity-capillary solution space at two values of the energy, which correspond to the energy of a gravity wave of approximately fifty and eighty percent of the amplitude of the limiting Stokes wave.
Surface tension enters our formulation via the (inverse) Bond number.
We restrict the Bond number to a range of small values, which for water corresponds to travelling wave solutions of wavelength 5cm to 9cm.
Solutions studied with the largest value of energy considered are all superharmonically unstable with the most unstable mode having with a very high temporal frequency.
At the relatively smaller value of energy, superharmonically stable solutions exist along portions of each solution branch. 
For these solutions there exist perturbation wavenumbers at which subharmonic instability emerges.
However, we find that stabilisation of the long-wave modulational instability occurs at much smaller values of the surface tension than predicted by weakly-nonlinear theory, and that even very small fluctuations in the surface tension can lead to significant changes in the stability properties of solutions. 

\subsection{Outline of our paper}
\noindent The time-dependent boundary-value problem is given in section~\ref{sec:mathform}, in which boundary-integral formulations are described for both general time-dependent solutions in section~\ref{sec:TDBIF} and for steadily travelling wave solutions in section~\ref{sec:steadyconf}.
The method used to study temporal stability of the travelling wave solutions is described in section~\ref{sec:linearstability}.
The subharmonic instabilities of surface gravity waves are described in section~\ref{sec:gravitystab}, and new results for the subharmonic instability of finite-amplitude gravity-capillary waves are given in section~\ref{sec:gcstab}.
A discussion of our results and our conclusions are offered in section~\ref{sec:disc}.

\section{Mathematical formulation}
\label{sec:mathform}
\noindent We consider a two-dimensional inviscid, irrotational and incompressible fluid that is bounded above by the periodic free surface $y=\zeta(x,t)$ and extends without bound as $y \to -\infty$.
The velocity field $(u,v)$ is expressed as the gradient of the velocity potential $\phi$ by writing $(u,v)=\nabla \phi$, for which the free-boundary problem of the Euler equations can be simplified to a potential-flow formulation that involves the solutions $\phi(x,y,t)$ and $\zeta(x,t)$. 
By introducing a comoving frame of speed $c$, we nondimensionalise length scales with respect to the wavelength $\lambda$, and velocity scales with respect to $c$ to obtain the nondimensional potential-flow formulation,
\begin{subequations}\label{eq:main}
\begin{align}
\label{eq:main1}
\phi_{xx} +\phi_{yy}=0& \qquad \text{for} ~~ y \leq \zeta(x,t),\\
\label{eq:main2}
\phi_t-\phi_{x}  +\frac{1}{2}(\phi_x^2+\phi_y^2) + \frac{\zeta}{F^2}-\frac{B}{F^2} \frac{\zeta_{xx}}{(1+\zeta_x^2)^{3/2}} =0& \qquad \text{at} ~~~ y=\zeta(x,t),\\
\label{eq:main3}
\zeta_t  = \phi_y+\zeta_x(1-\phi_x) & \qquad \text{at} ~~~ y=\zeta(x,t),\\
\label{eq:main4}
\phi_{x} \to 0, \quad \phi_{y} \to 0& \qquad \text{as} ~~ y \to -\infty.
\end{align}
\end{subequations}
The boundary-value problem consists of Laplace's equation \eqref{eq:main1}, with dynamic and kinematic boundary conditions (\hyperref[eq:main2]{\ref{eq:main2},c}) at the surface, and decay conditions \eqref{eq:main4} at infinite depth.
The two nondimensional constants appearing in \eqref{eq:main2} are the Froude and Bond numbers, defined by
\begin{equation}\label{eq:nondimparams}
F= \frac{c}{\sqrt{g \lambda}} \quad \text{and} \quad B=\frac{\sigma}{\rho g \lambda^2}
\end{equation}
where $c$ is the speed of the comoving frame, $g$ is the constant acceleration due to gravity, $\lambda$ is the wavelength, $\sigma$ is the constant coefficient of surface tension, and $\rho$ is the fluid density.

The total energy of the system, given by
\begin{equation}\label{eq:energy}
\mathcal{E}= \int_{-1/2}^{1/2}\bigg[\frac{F^2}{2}\phi (\phi_y-\phi_x \zeta_x)\rvert_{_{y=\zeta}} +B(\sqrt{1+\zeta_x^2}-1)+\frac{\zeta^2}{2}\bigg] \mathrm{d}x,    
\end{equation}
is a Hamiltonian and time-conserved quantity of the system \cite{zakharov1968stability}, which consists of kinetic, capillary potential, and gravitational potential energies.
To express the kinetic energy in \eqref{eq:energy} as an integral over the surface, $y=\zeta(x,t)$, Green's first identity was applied to the two-dimensional integral of $u^2 + v^2$.
The value of the energy will be fixed as an amplitude parameter when solving for steady solutions in the $(B,F)$ bifurcation space.

\subsection{The time-dependent boundary-integral formulation}
\label{sec:TDBIF}
\noindent We consider a time-dependent conformal map that takes the two-dimensional free-boundary problem \eqref{eq:main} to a one-dimensional boundary-integral problem.
This time-dependent conformal mapping for free-surface waves was developed by
Dyachenko \emph{et al.} \cite{dyachenko1996nonlinear}, and further details on the derivation are given by Choi \& Camassa \cite{choi1999exact} in a fluid of finite depth.
The conformal map takes the fluid domain in the $(x,y)$ plane to the lower-half $(\xi,\eta)$ conformal domain, for which a two-dimensional boundary-value problem is obtained that involves kinematic and dynamic boundary conditions evaluated at $\eta=0$.
By writing $x=x(\xi,\eta,t)$ and $y=y(\xi,\eta,t)$, further simplifications are made to the system by first defining the surface solutions
\begin{equation*}
\begin{aligned}\label{eq:Tmapvariables}
X(\xi,t)=x(\xi,0,t),& \qquad \Phi(\xi,t)=\phi(x(\xi,0,t),y(\xi,0,t),t),\\
Y(\xi,t)=\zeta(x(\xi,0,t),t),& \qquad \Psi(\xi,t) = \psi(x(\xi,0,t),y(\xi,0,t),t),
\end{aligned}
\end{equation*}
where $\psi$ is the harmonic conjugate of $\phi$.
A boundary-integral formulation is then obtained for these surface solutions, which consists of time-evolution equations for $Y$ and $\Phi$,
\begin{subequations}\label{eq:mapevolve}
\begin{align}\label{eq:Yevolve}
Y_t=\frac{X_{\xi}(Y_{\xi}-\Psi_{\xi})}{J}-Y_{\xi} \mathcal{H} \bigg[\frac{Y_{\xi}-\Psi_{\xi}}{J}\bigg],\\
\label{eq:Phievolve}
\Phi_t=\frac{\Psi_{\xi}^2-\Phi_{\xi}^2}{2J} - \frac{Y}{F^2}+ \frac{B}{F^2}\frac{X_{\xi}Y_{\xi \xi}-Y_{\xi}X_{\xi \xi}}{J^{3/2}}+\frac{X_{\xi}\Phi_{\xi}}{J}-\Phi_{\xi}\mathcal{H} \bigg[\frac{Y_{\xi}-\Psi_{\xi}}{J}\bigg],
\end{align}
and harmonic relations
\refstepcounter{equation}\label{eq:harmonic1}
\refstepcounter{equation}\label{eq:harmonic2}
\begin{equation}
X_{\xi}=1-\mathcal{H}[Y_{\xi}] \quad \text{and} \quad \Psi = \mathcal{H}[\Phi].
\tag{\ref*{eq:harmonic1},d}
\end{equation}
\end{subequations}
Here, $\mathcal{H}[Y]=\int_{-1/2}^{1/2}Y(\xi^{'}) \cot{[\pi(\xi^{'}-\xi)]}\mathrm{d}\xi^{'}$ is the periodic Hilbert transform, and $J=X_{\xi}^2+Y_{\xi}^2$ is the Jacobian of the conformal map.
Under the conformal mapping, the energy expression \eqref{eq:energy} simplifies to
\begin{equation}\label{eq:energymap}
\mathcal{E}= \int_{-1/2}^{1/2}\bigg[\frac{F^2}{2} \Psi_{\xi} \Phi +B(\sqrt{J}-X_{\xi})+\frac{Y^2X_{\xi}}{2}\bigg] \mathrm{d} \xi.    
\end{equation}

\subsection{The time-independent boundary-integral formulation}
\label{sec:steadyconf}
\noindent Solutions which are steady in the comoving frame (corresponding to permanently propagating waves in the lab frame) have $Y_t=0$ and $\Phi_t=0$ in the boundary-integral system \eqref{eq:mapevolve}.
The stability of these with respect to subharmonic perturbations is the focus of this work.
Steady solutions have $\Psi=Y$, $\Phi=X-\xi$, and satisfy the equation
\begin{equation}\label{eq:steadybern}
\frac{1}{2}\bigg(1-\frac{1}{J} \bigg) - \frac{Y}{F^2}+ \frac{B}{F^2}\frac{X_{\xi}Y_{\xi \xi}-Y_{\xi}X_{\xi \xi}}{J^{3/2}}=0
\end{equation}
where $X$ and $Y$ are related by the Hilbert transform via \eqref{eq:harmonic1}.
Each solution of this steady problem has a corresponding value of $B$ and $F$.
Rather than pick values for both $B$ and $F$, and then attempt to obtain a solution, we will instead enforce an amplitude condition for which one of $B$ or $F$ is considered an unknown.
Two different amplitude conditions will be used in this work which are the wave elevation $\mathcal{A}$, defined by
\begin{equation}
\label{eq:amp}
\mathcal{A}=\frac{1}{2}\big[y(0)-y(1/2)\big],\\
\end{equation}
and the energy $\mathcal{E}$ from \eqref{eq:energymap}.
The wave amplitude $\mathcal{A}$ is a very good measure of nonlinearity for gravity waves, as solutions in the small-amplitude regime, with $0<\mathcal{A} \ll 1$, and the steepest wave of limiting form, with $\mathcal{A}\approx 0.0705$, are connected by solutions with strictly increasing values of $\mathcal{A}$ \cite{cokelet1977steep}.
In the computation of periodic gravity-capillary waves, various amplitude parameters have been employed. These include the amplitude $\mathcal{A}$ by Schwartz \& Vanden-Broeck \cite{schwartz1979numerical}, the Fourier coefficient of $\cos{(2 \pi x)}$ by Chen \& Saffman \cite{chen1980steady}, and the energy $\mathcal{E}$ by Shelton \emph{et al.} \cite{shelton2021structure}.
In this latter study, it was demonstrated numerically that for fixed energy, a countably-infinite number of self-similar solutions branches emerge in the small-surface-tension limit.
Analogous solution branches were also calculated by Champneys \emph{et al.}~\cite{champneys2002true} for fixed Froude number, $F$.
We will obtain solutions to this system by applying Newton iteration to a spectral numerical scheme, and further details of this method are given in Appendix~\ref{app:steady}.

\subsection{Linear stability}
\label{sec:linearstability}
\noindent The methodology used to study the stability of subharmonic perturbations to steady solutions of the gravity-capillary wave problem is now introduced.
First, we consider time-dependent perturbations about steady solutions, $X_0(\xi)$, $Y_0(\xi)$, $\Phi_0(\xi)$, and $\Psi_0(\xi)$, of the conformal formulation from section~\ref{sec:steadyconf} by writing
\begin{equation}\label{eq:expansion}
\{ X, Y, \Phi, \Psi \} \sim \{ X_0, Y_0, \Phi_0, \Psi_0 \} + \epsilon \{ X_1, Y_1, \Phi_1, \Psi_1 \}+O(\epsilon^2).
\end{equation}
Here, $X_1(\xi,t)$, $Y_1(\xi,t)$, $\Phi_1(\xi,t)$, and $\Psi_1(\xi,t)$, are the time-dependent perturbation quantities.
Equations for these perturbation quantities are obtained at $O(\epsilon)$ after substituting expansions \eqref{eq:expansion} into the time-dependent governing system (\hyperref[eq:Yevolve]{\ref{eq:Yevolve}--d}), which yields
\begin{subequations}\label{eq:mapevolveOep}
\begin{align}\label{eq:YevolveOep}
&Y_{1t}=\frac{X_{0\xi}(Y_{1\xi}-\Psi_{1\xi})}{J_0}-Y_{0\xi} \mathcal{H} \bigg[\frac{Y_{1\xi}-\Psi_{1\xi}}{J_0}\bigg],\\
\label{eq:PhievolveOep}
&\begin{aligned}
\Phi_{1t}&=\frac{\Psi_{0 \xi}\Psi_{1 \xi}-\Phi_{0 \xi} \Phi_{1 \xi}+X_{0 \xi} \Phi_{1 \xi} + \Phi_{0 \xi} X_{1 \xi}}{J_0}-\frac{(\Psi_{0 \xi}^2 - \Phi_{0 \xi}^2 +2X_{0 \xi} \Phi_{0 \xi})(X_{0 \xi}X_{1 \xi}+Y_{0 \xi}Y_{1 \xi})}{J_0^2}\\
&\hphantom{=}~ - \frac{Y_1}{F^2}+ \frac{B}{F^2}\frac{X_{0\xi}Y_{1\xi \xi}+Y_{0\xi \xi}X_{1\xi}-Y_{0\xi}X_{1\xi \xi}-X_{0\xi \xi}Y_{1\xi}}{J_0^{3/2}}\\
&\hphantom{=}~ -\frac{3B}{F^2}\frac{(X_{0\xi}Y_{0\xi \xi}-Y_{0\xi}X_{0\xi \xi})(X_{0 \xi}X_{1 \xi}+Y_{0 \xi}Y_{1 \xi})}{J_0^{5/2}}-\Phi_{0\xi}\mathcal{H} \bigg[\frac{Y_{1\xi}-\Psi_{1\xi}}{J_0}\bigg],
\end{aligned}
\end{align}
\refstepcounter{equation}\label{eq:harmonic1Oep}
\refstepcounter{equation}\label{eq:harmonic2Oep}
\begin{equation}
X_{1\xi}=-\mathcal{H}[Y_{1\xi}], \qquad  \Psi_1 = \mathcal{H}[\Phi_1].
\tag{\ref*{eq:harmonic1Oep},d}
\end{equation}
\end{subequations}

Crucially, the above equations governing the perturbation quantities are linear but with complicated coefficients depending on the leading-order steady solutions that are only known numerically.
By applying separation of variables we then see that each solution may be written as
\begin{equation}\label{eq:floquetansatz}
\{ X_1,Y_1,\Phi_1,\Psi_1 \}=\mathrm{e}^{\sigma t} \sum_{n=-\infty}^{\infty} \{a_n,b_n,c_n,d_n \} \mathrm{e}^{2 (n+p) \pi \mathrm{i} \xi},
\end{equation}
where the complex-valued constant $\sigma$ is unknown, and the real-valued constant $p$, the Floquet exponent, is specified.

Due to the time-dependent formulation of the gravity-capillary wave problem having an equivalent Hamiltonian formulation, there is four-fold symmetry in the eigenvalue spectrum. That is, if there exists a solution with a certain value of $(\sigma,p)$, three additional solutions are obtained with $(-\sigma,-p)$, $(\sigma^{*},-p)$, and $(-\sigma^{*},p)$, where $\sigma^*$ is the complex conjugate of $\sigma$. 
Therefore, we will restrict all results to the quadrant $\text{Re}[\sigma] \geq 0$ and $\text{Im}[\sigma] \geq 0$, as well as the range $0 \leq p \leq 1/2$.
Details on our numerical implementation of this linear stability problem are given in Appendix~\ref{app:linear}, where the system \eqref{eq:mapevolveOep} is converted into a generalised eigenvalue problem for the eigenvector $\{a_{-N}, \ldots a_{N}, c_{-N}, \ldots, c_{N} \}$ and eigenvalue $\sigma$.

The real part of $\sigma$ in solution \eqref{eq:floquetansatz} corresponds to the growth rate of the perturbation, and the imaginary part to the temporal oscillation.
The value of $\sigma$ will depend on the choice of the Floquet exponent $p$.
Therefore, the growth rate of the perturbation will depend on the associated perturbation wavelength, $1/p$.
\emph{Superharmonic} perturbations, which have the same wavelength as the leading-order steady solution, are obtained with $p=0$.
\emph{Subharmonic} perturbations, which have a different wavelength compared to the steady solution, are obtained with $0<p \leq 1/2$.
A solution is said to be unstable if there exists an eigenvalue $\sigma$ with positive real part, and stable if every eigenvalue satisfies $\text{Re}[\sigma] \leq 0$.
Due to the four-fold symmetry of the eigenvalue spectrum, we strengthen the condition of stability to be the criteria that every eigenvalue satisfies $\text{Re}[\sigma] = 0$.
Therefore, a solution is \emph{superharmonically stable} if every eigenvalue satisfies $\text{Re}[\sigma] = 0$ for $p=0$, and \emph{superharmonically unstable} if there exists an eigenvalue with $\text{Re}[\sigma] \neq 0$ when $p=0$.
Furthermore, a solution is \emph{subharmonically stable} if every eigenvalue satisfies $\text{Re}[\sigma] = 0$ for all $0<p \leq 1/2$, and \emph{subharmonically unstable} if for any $0<p \leq 1/2$ there exists an eigenvalue with $\text{Re}[\sigma] \neq 0$.
Additionally, the modulational instability (also known as the Benjamin-Feir instability) is an instability with respect to long wave perturbations and so occurs in the limit of $p \to 0$.
By expanding about $\sigma=0$ for small $p$ as $\sigma \sim \sigma_1 p +O(p^2)$, a solution is said to be modulationally unstable if $\sigma_1 \neq 0$, and modulationally stable otherwise.

\section{Subharmonic instability of gravity waves}
\label{sec:gravitystab}
\begin{figure}
\centering
\includegraphics[scale=1]{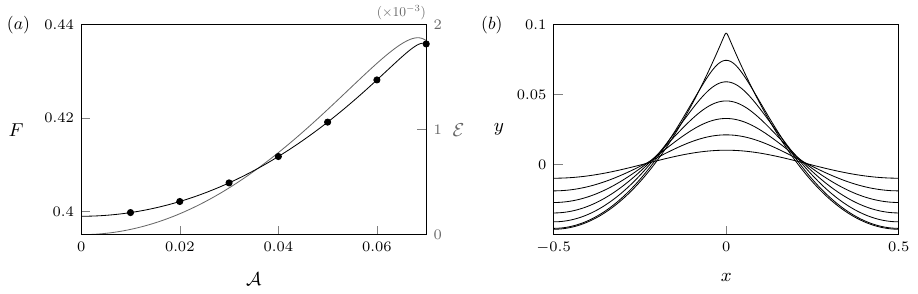}
\caption{\label{fig:gravsols} Solutions are shown for travelling surface gravity waves on infinite depth, which satisfy equation \eqref{eq:steadybern} with zero Bond number.
In panel $(a)$, the dispersion relation (black line) is shown for different values of the amplitude $\mathcal{A}$ defined in \eqref{eq:amp}. The associated energy $\mathcal{E}$ from \eqref{eq:energymap} is also shown (grey line) for each value of $\mathcal{A}$. Seven solution profiles are shown in panel $(b)$ for $\mathcal{A}= \{0.01,0.02,0.03,0.04,0.05,0.06,0.07 \}$, which have energy values $\mathcal{E}= \{0.00004995,0.0001991,0.0004450,0.0007812,0.001192,0.001632,0.001849\}$. These solutions were obtained using the numerical scheme detailed in Appendix~\ref{app:steady} with $N=4096$ Fourier coefficients.}
\end{figure}

\noindent In this section we summarise known results for the subharmonic instability of travelling gravity waves on a fluid of infinite depth. 
The solution space for steadily travelling gravity waves is shown in figure~\ref{fig:gravsols}.
When the amplitude $\mathcal{A}$ is small, solutions are described by the linear approximation $F \sim (2 \pi)^{-1/2}+O(\mathcal{A})$ and $\zeta(x) \sim \mathcal{A}\cos{(2 \pi x)}+O(\mathcal{A}^{2})$.
As the amplitude increases, the solution profile approaches the limiting Stokes wave with an amplitude of $\mathcal{A} \approx 0.0705$ and an energy of $\mathcal{E} \approx 0.00184$ \cite{cokelet1977steep}. 
Values for the Froude number $F$ and energy $\mathcal{E}$ are shown in figure~\ref{fig:gravsols}$(a)$ for different values of the amplitude, and corresponding solution profiles are shown in $(b)$.

It was shown by Longuet-Higgins \cite{longuet1978instabilities} that gravity waves are stable to superharmonic perturbations for amplitude values $0 \leq \mathcal{A}  \lessapprox 0.069$.
Beyond this critical amplitude, corresponding to the maximum value of the Energy $\mathcal{E}$ in figure~\ref{fig:gravsols}(a) \cite{tanaka1983stability}, it was shown that superharmonic instability emerges for which the time-dependent perturbations have a high spatial frequency. 
Gravity waves of all amplitude are known to be unstable with respect to subharmonic perturbations.
When the amplitude is small, this instability takes the form of a long-wave perturbation as famously shown by Benjamin \& Feir \cite{benjamin1967disintegration}.
At finite values of the amplitude, Longuet-Higgins \cite{10.1098/rspa.1978.0081} calculated growth rates associated with certain subharmonic perturbations, and showed that the instability at certain frequencies was stabilised beyond a certain value of $\mathcal{A}$.
More specifically, the long-wave modulational instability is stabilised above the value $\mathcal{A}\approx 0.0545$ \cite{deconinck2011instability}.

\begin{figure}
\centering
\includegraphics[scale=1]{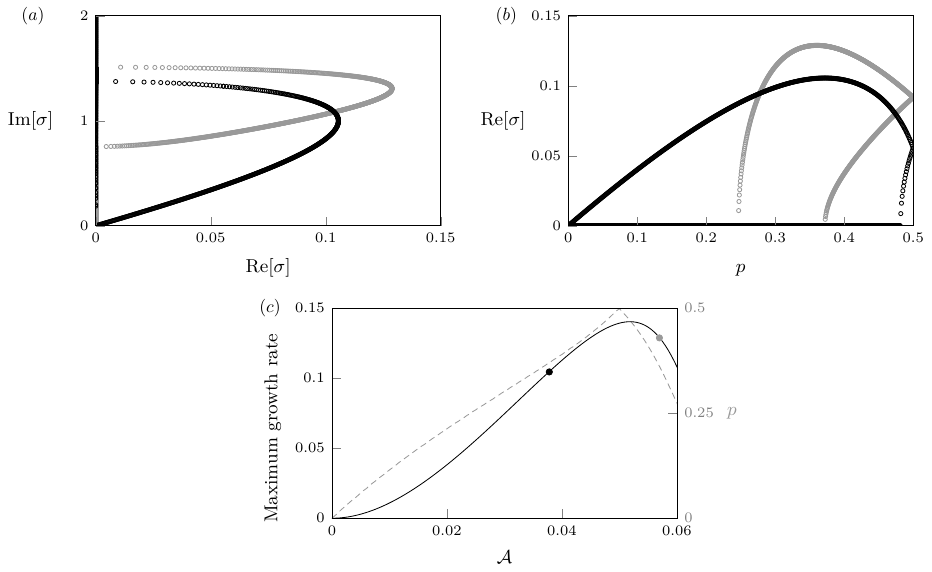}
\caption{\label{fig:gravstab} Subharmonic stability results are shown for gravity waves on infinite depth. In panels $(a,b)$, the growth rates Re$[\sigma]$ and frequencies Im$[\sigma]$ are shown for linear subharmonic perturbations to two large-amplitude solutions: one with $\mathcal{E}=0.0007$ (black) and another with $\mathcal{E}=0.0015$ (grey). Due to symmetry, only the quadrant Re$[\sigma]>0$ and Im$[\sigma]>0$ is shown in panel $(a)$. The corresponding value of the Floquet exponent $p$ is shown for each growth rate in panel $(b)$. Recall that the wavelength of the perturbation is given by $1/p$. In $(c)$, we show the largest growth rate (black) that exists for each amplitude value $\mathcal{A}$, as well as the corresponding value of $p$ (dashed grey) at which this maximum occurs. The two marked locations correspond to the maximum growth rate from each solution studied in $(a,b)$. These results were obtained using the numerical scheme described in Appendix~\ref{app:linear} applied to $501$ equally spaced values of $0 \leq p \leq 0.5$.}
\end{figure}
Results for the subharmonic instability of gravity waves are shown in figure~\ref{fig:gravstab}. In panel $(c)$, the maximum growth rate is shown for a range of amplitude values, and the associated value of the Floquet exponent $p$ for each of these maximum growth rates is shown dashed.
All of these solutions are subharmonically unstable due to a positive growth rate existing for $0<p \leq 1/2$.
We see that when $\mathcal{A}$ is small, the long-wave modulational instability is dominant (since $p \to 0$ and $\mathcal{A} \to 0$).
As the amplitude increases, the wavelength $1/p$ associated with the largest growth rate decreases.
The two marked locations correspond to solutions with $\mathcal{E}=0.0007$ (black) and $\mathcal{E}=0.0015$ (grey), for which the full spectrum of instability is shown in $(a,b)$.
In panel $(a)$, all eigenvalues are shown in the first quadrant.
For each value of $p$ a discrete set of eigenvalues is obtained, and as $p$ changes value these eigenvalues trace out a continuous curve in the quadrant.
These results were obtained by solving the linear stability problem with $501$ different values of $0 \leq p \leq 0.5$.
In figure~\ref{fig:gravstab}$(b)$, we show the growth rates Re$[\sigma]$ associated with each value of $p$.
Both solutions are superharmonically stable, as there are no non-zero growth rates when $p=0$.
The solution with $\mathcal{E}=0.0007$ is modulationally unstable as there exist positive growth rates in the long-wave limit of $p \to 0$, whereas that with $\mathcal{E}=0.0015$ is modulationally stable as all growth rates are zero as $p \to 0$.
Nevertheless, the solution with $\mathcal{E}=0.0015$ is still subharmonically unstable with respect to perturbations that have $0.247 < p \leq 0.5$.
These two solutions, obtained with $B=0$, will form a comparison of our stability investigations when surface tension is included in the steady formulation for the travelling wave.

\section{Subharmonic instability of gravity-capillary waves}
\label{sec:gcstab}
\noindent We now consider the stability of travelling wave solutions subject to the effects of both gravity and surface tension. 
Firstly, in section~\ref{sec:weaklynonlin} we use weakly-nonlinear theory to analytically derive the stability of long-wave perturbations to small-amplitude solutions. Branches of fully nonlinear solutions are then considered in section~\ref{sec:gcsols}, and the stability of these studied in section~\ref{sec:gcstab1} for energy $\mathcal{E}=0.0007$ and section~\ref{sec:gcstab2} for $\mathcal{E}=0.0015$.

\subsection{Modulational instability of weakly-nonlinear solutions}
\label{sec:weaklynonlin}
\noindent We now examine the modulational instability of gravity-capillary waves predicted by weakly-nonlinear theory. 
By perturbing about the exact solution $Y(\xi)=0$ of the steady system \eqref{eq:steadybern} as $Y \sim \epsilon Y_1 +\cdots$, the linear solution and dispersion relation are found to be
\begin{subequations}
\refstepcounter{equation}\label{eq:linearsol}
\refstepcounter{equation}\label{eq:lineardisp}
\begin{equation}
Y_1(\xi)=A \mathrm{e}^{2 k \pi \mathrm{i} \xi} + c.c., \qquad F=\sqrt{\frac{1+4 k^2\pi^2 B}{2k\pi}},
\tag{\ref*{eq:linearsol},b}
\end{equation}
\end{subequations}
where $A$ is a constant, $k$ is the wavenumber, and $c.c.$ denotes complex conjugate. 
The development of this linear solution over large length- and time-scales can be described by a multiple-scales analysis, performed on the time-dependent system \eqref{eq:mapevolve} up to $O(\epsilon^3)$ in the small-amplitude expansion, which yields a nonlinear-Schrödinger (NLS) equation \cite{hogan1985fourth,dias1999nonlinear} for the amplitude $A$ associated with solution \eqref{eq:linearsol}.
Expressed in our nondimensional formulation, this evolution equation is
\begin{subequations}
\begin{equation}
\label{eq:GCNLS}
\mathrm{i}A_{\tau} + \alpha A_{\sigma \sigma} = \beta |A|^2 A,
\end{equation}
where the coefficients $\alpha$ and $\beta$ are given by
\begin{equation}
\label{eq:GCNLScoeff}
\alpha=\frac{48 \pi^4 B^2+24\pi^2 B-1}{8(2 \pi)^2(1+ 4 \pi^2 B)^2} ,\qquad \beta=\pi^2\frac{8 \pi^4 B^2 +  \pi^2 B + 2}{(1-8\pi^2B)(1+4 \pi^2B)}.
\end{equation}
\end{subequations}
In the NLS equation \eqref{eq:GCNLS}, $\tau=\epsilon^2 t$ is the $O(\epsilon^2)$ timescale and $\sigma=\epsilon(x+ [1-c_g] t)$ represents the difference between a frame of reference moving with the nondimensional group velocity, $c_g$, and another with unit nondimensional speed.
Note that the values of the coefficients of the NLS equation in \eqref{eq:GCNLScoeff} differ to \emph{e.g.} Dias \& Kharif \cite{dias1999nonlinear} due to our formulation being nondimensionalised with respect to the wavelength $\lambda$, rather than the wavenumber $k$, which results in factors of $\lambda/k=2 \pi$ emerging throughout.

Modulational stability of the linear monochromatic wave from \eqref{eq:linearsol} can be immediately inferred from the sign of $\beta/\alpha$.
This is achieved by considering the stability of the exact solution $A(\tau,\sigma)=a\mathrm{e}^{-\mathrm{i} \beta a^2 \tau}$ of \eqref{eq:GCNLS} with respect to long wave perturbations \cite{trichtchenko2019stability}, which are stable when $\beta/\alpha >0$ (known as the defocusing case) and unstable when $\beta/\alpha<0$ (known as the focusing case).
We see from the coefficient values in \eqref{eq:GCNLScoeff} that there is a change in sign of $\beta/\alpha$ at the two values
\begin{equation}
\label{eq:signvals}
B_1=\frac{2\sqrt{3}-3}{12 \pi^2}, \qquad B_2=\frac{1}{8 \pi^2},
\end{equation}
where $B_1 < B_2$ with $B_1 \approx 0.003919$, $B_2 \approx 0.01267$.
Note that there are resonance conditions whenever two linear solutions with different wavenumbers have the same value of the Froude number in \eqref{eq:lineardisp}.
This occurs at $B=(4 \pi^2 k)^{-1}$, and so the value of $B_2$ in \eqref{eq:signvals} is identified as a resonance between linear waves with wavenumbers one and two.
The NLS equation \eqref{eq:GCNLS} is only valid away from these resonant values of the Bond number.

Small-amplitude gravity waves, which have $B=0$, are known to be modulationally unstable \cite{benney1967propagation}. 
We confirm this result since $\beta/\alpha<0$ when $B=0$.
The weakly-nonlinear theory of this section has shown that this instability changes with the inclusion of surface tension.
When $0 \leq B < B_1$, we have modulational instability.
This changes to modulational stability for $B_1<B<B_2$, and finally reverts back to modulational instability when $B_2<B$.
Note that while we derived these results using formal asymptotics in the small-amplitude limit, they have been proven by both Hur \& Yang~\cite{hur2023unstable} and Sun \& Wahl{\'e}n~\cite{sun2025spectral}.

\subsection{The gravity-capillary wave solution space}
\label{sec:gcsols}
\noindent Travelling gravity-capillary wave solutions exist within a three-dimensional parameter space characterised by the Bond number, $B$, the Froude number, $F$, and an amplitude parameter, which we specify to be the energy $\mathcal{E}$ defined in \eqref{eq:energymap}. These solutions satisfy the boundary-integral system, consisting of the integral equation \eqref{eq:harmonic1} and the nonlinear differential equation \eqref{eq:steadybern}.

Fully nonlinear travelling-wave solutions to the gravity-capillary wave problem were first obtained numerically by Schwartz \& Vanden-Broeck \cite{schwartz1979numerical}, who demonstrated the existence of different classes of solutions that exist in the regimes of large and small surface tension. When the surface tension was large, overturning solutions were obtained that closely resembled the pure-capillary analytical solutions of Crapper \cite{crapper1957exact}.
At smaller values of the surface tension, they obtained solution profiles that closely resembled Stokes waves (which are solutions of the system with $B=0$) containing small-amplitude capillary ripples on the wave surface. Analogous solutions were later investigated in detail by Champneys \emph{et al.} \cite{champneys2002true}, who found that all solutions obtained numerically with small non-zero values of surface tension contained these oscillatory capillary modes.

\begin{figure}
\centering
\includegraphics[scale=1]{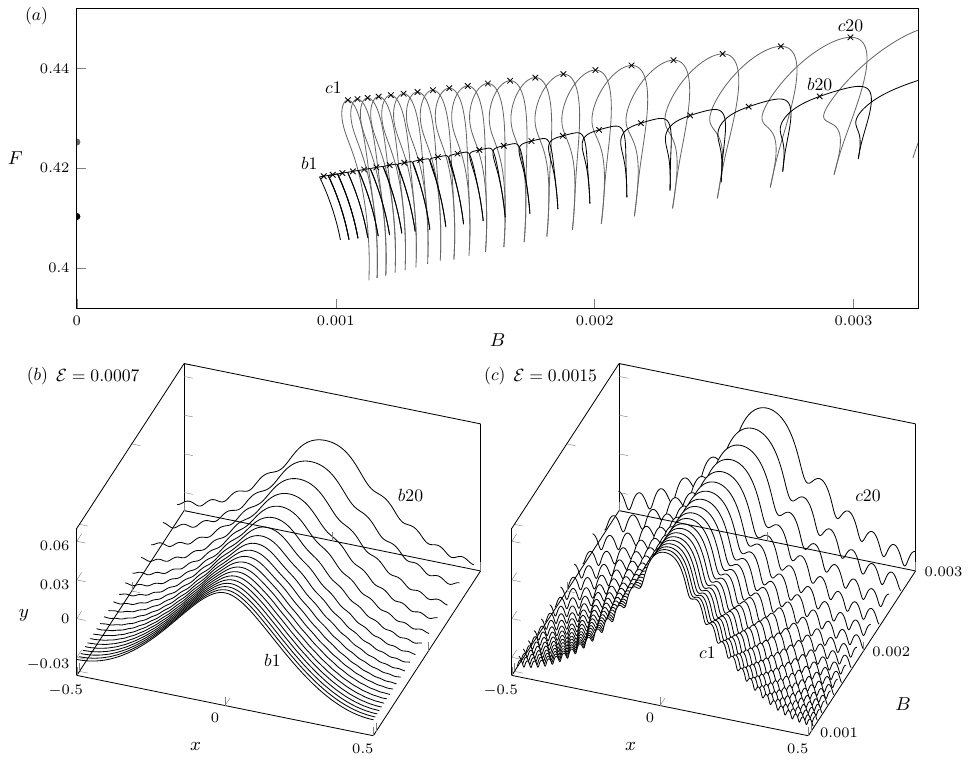}
\caption{\label{fig:gcsolspace} Results for travelling gravity-capillary waves of large amplitude are shown for the two energy levels $\mathcal{E}=0.0007$ and $\mathcal{E}=0.0015$. Panel $(a)$ shows values of the Bond, $B$, and Froude, $F$, numbers, defined in \eqref{eq:nondimparams}, at which solutions exist. Twenty branches have been calculated within the range $0.001<B<0.003$ for each of the two energy values. One solution from each branch with $\mathcal{E}=0.0007$ and $\mathcal{E}=0.0015$ is shown in panels $(b)$ and $(c)$, respectively. These solutions were obtained using the numerical scheme from Appendix~\ref{app:steady} with $N=2047$ Fourier coefficients.}
\end{figure}
Branches of nonlinear solutions to the gravity-capillary wave problem were studied in detail by Shelton \emph{et al.} \cite{shelton2021structure} by fixing the energy $\mathcal{E}$ as an amplitude condition.
It was demonstrated that a countably infinite set of branches emerged in the small-surface-tension limit such that along each branch, denoted by $G_{n \rightarrow n+1}$, the wavenumber of the capillary modes transitioned from $n$ to $n+1$. 
These branches each bifurcated from additional branches with fundamental frequencies $n$ and $n+1$.
Twenty of these branches are shown in figure~\ref{fig:gcsolspace}$(a)$ for both $\mathcal{E}=0.0007$ and $\mathcal{E}=0.0015$.
The branches calculated with $\mathcal{E}=0.0007$, which corresponds to the energy of a Stokes wave of approximately half the limiting amplitude, are those calculated by \cite{shelton2021structure}, and solution profiles on each of the branches are shown in figure~\ref{fig:gcsolspace}$(b)$.
These solutions were chosen to ensure that they are horizontally half way between the two locations on the same branch with infinite and zero gradient. 
The displayed solutions are steep travelling waves which contain oscillatory capillary ripples.
The bifurcation branches calculated with $\mathcal{E}=0.0015$, corresponding to the energy of a Stokes wave of approximately eighty percent of the limiting amplitude, are also shown in figure~\ref{fig:gcsolspace}$(a)$.
These solutions are more energetic than those calculated by \cite{shelton2021structure}, and solutions on each of the twenty branches are shown in \ref{fig:gcsolspace}$(c)$.
The capillary oscillations are more prominent in these displayed solutions.

\begin{figure}
\centering
\includegraphics[scale=1]{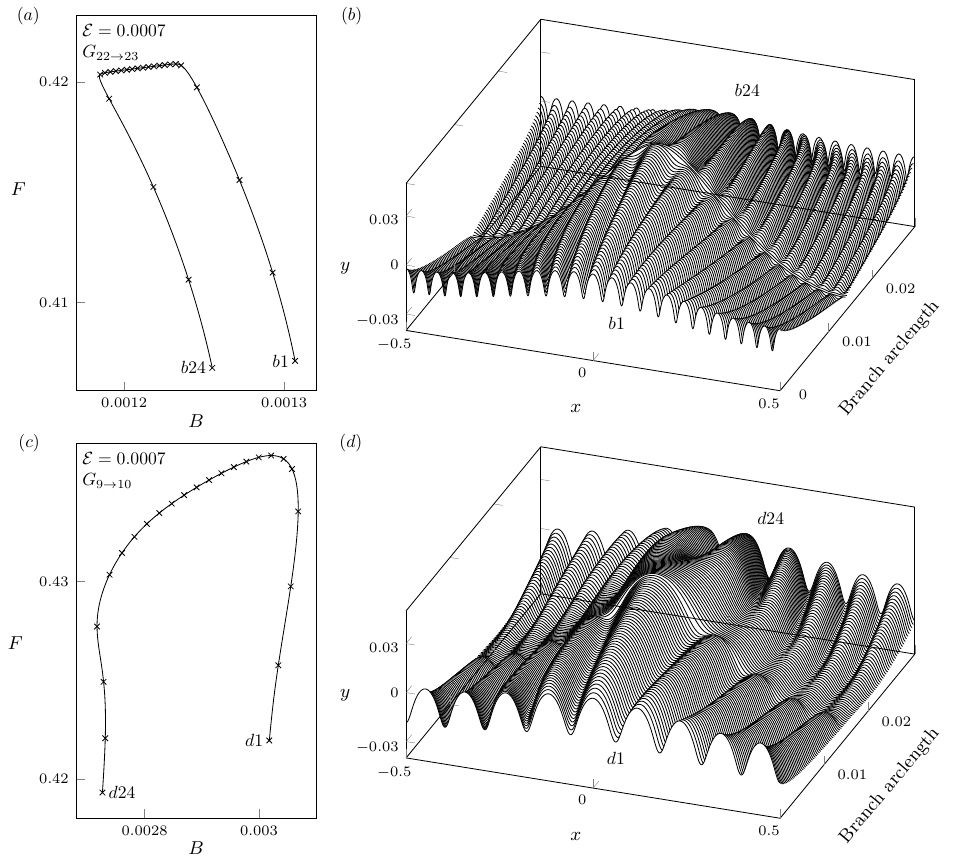}
\caption{\label{fig:branches0007} The behaviour of solutions along two branches from figure~\ref{fig:gcsolspace}$(a)$ with $\mathcal{E}=0.0007$ is shown. The branch $G_{22 \rightarrow 23}$ is shown in panel $(a)$. This branch connects two (omitted) branches each with a fundamental frequency of $22$ and $23$, respectively. The branch $G_{9 \rightarrow 10}$ is shown in panel $(c)$. The marked locations along both of these two branches denote the twenty four solutions used to study temporal stability later in figures~\ref{fig:finger1stab} and \ref{fig:finger2stab}. The behaviour of solution profiles along both these branches is shown in panels $(b,d)$. In these, one hundred profiles are displayed for which each was chosen to have uniform spacing in arclength along the respective branch. These solutions were obtained with $N=1023$ Fourier modes using the numerical scheme described in Appendix~\ref{app:steady}.}
\end{figure}

The behaviour of solutions along two branches with $\mathcal{E}=0.0007$ from figure~\ref{fig:gcsolspace}$(a)$ is shown in figure~\ref{fig:branches0007}.
The branch $G_{22 \rightarrow 23}$ is shown in \ref{fig:branches0007}$(a)$, and branch $G_{9 \rightarrow 10}$ is shown in \ref{fig:branches0007}$(c)$.
Solution profiles along these branches are shown in figure~\ref{fig:branches0007}$(b,c)$.
At the bottom right of the branch $G_{n \rightarrow n+1}$, solutions are dominated by the mode with wavenumber $n$.
Moving along the respective branch, the primary mode grows in the solution and the amplitude of the oscillatory mode decreases.
Near the top of the solution branch, the solution attains its maximum amplitude and consists of a large-amplitude Stokes wave with small-amplitude capillary modes.
Then, along the end of the branch, the solution becomes dominated by a mode with wavenumber $n+1$.
It is through this process that the wavenumber of the capillary mode tends to infinity in the limit of small surface tension.
The energy $\mathcal{E}$ defined in \eqref{eq:energymap} has three components: kinetic, gravitational, and capillary potential energies.
The capillary energy is largest at either end of the branch (such that in the limit of small surface tension there exist solutions dominated by capillarity), and smallest along the top region of the branch where the capillary modes have the minimum amplitude.
\begin{figure}
\centering
\includegraphics[scale=1]{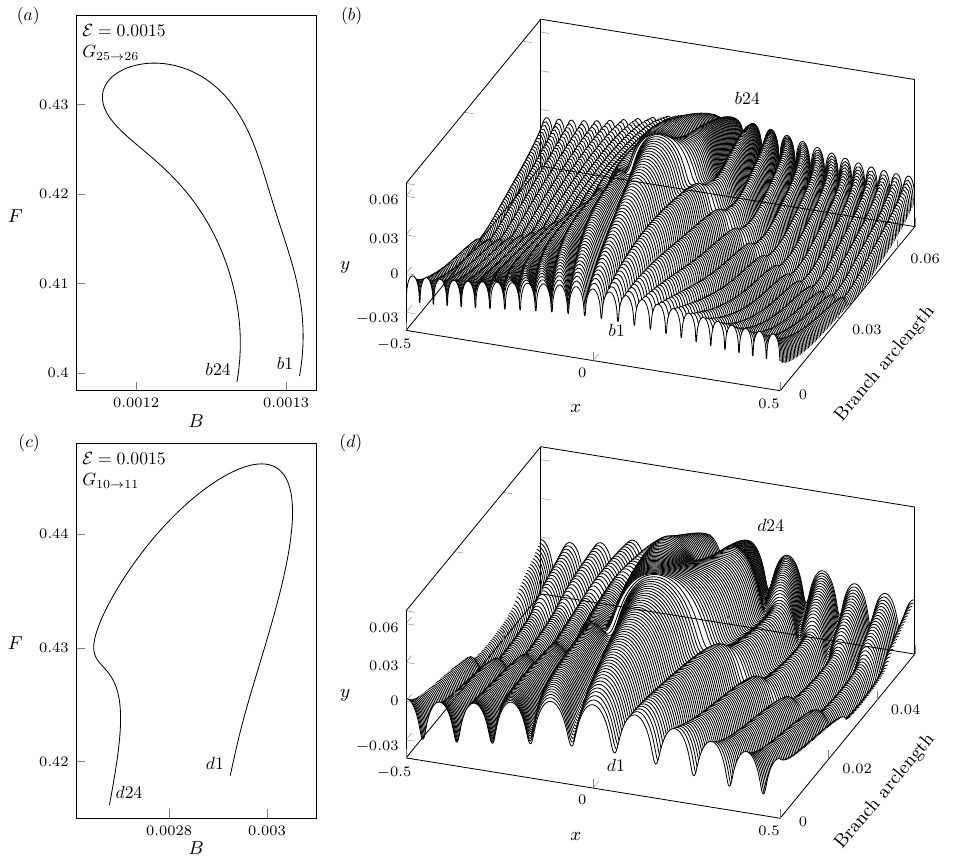}
\caption{\label{fig:branches0015} The behaviour of solutions along two branches from figure~\ref{fig:gcsolspace}$(a)$ with $\mathcal{E}=0.0015$ is shown. The branch $G_{25 \rightarrow 26}$ is shown in panel $(a)$, and the solutions along this branch shown in panel $(b)$. The branch $G_{10 \rightarrow 11}$ is shown in panel $(c)$, and solutions along the branch shown in panel $(d)$. The solutions in $(b,d)$ have equidistant spacing in arclength along the respective branch, and were calculated with $N=2047$ Fourier modes using the numerical scheme described in Appendix~\ref{app:steady}.}
\end{figure}
Solutions with $\mathcal{E}=0.0015$ along the two branches $G_{25 \rightarrow 26}$ and $G_{10 \rightarrow 11}$ are shown in figure~\ref{fig:branches0015}.
The behaviour is analogous to that which occurs with $\mathcal{E}=0.0007$, except for the ratio between the amplitude of the capillary mode and the gravity mode is disproportionately larger.
Along each branch, the extra wavenumber in the capillary mode (\emph{e.g.} from $n=22$ to $n=23$) is produced across a small portion of the branch close to where the Froude number is largest.

The behaviour of solutions along each branch can also be interpreted under the asymptotic limit of $B \to 0$.
The solutions along the sides of each branch are described by fully nonlinear multiple-scales theory, in which the fast scale has nonlinear dependence on the slow scale, and the dependency takes the form of Crappers exact solution \cite{crapper1957exact}. Resolving this multiple-scales regime remains an open problem, and requires the application of Kuzmak's fully nonlinear multiple-scales theory~\cite{kuzmak1959asymptotic,chapman2017analysis}.
Along the top of each solution branch where the capillary ripple amplitude is smallest, the asymptotic behaviour was resolved by Shelton \& Trinh~\cite{shelton2022exponential,shelton2024model}.
Here, the ripple amplitude is exponentially subdominant in the small-surface-tension limit and required an application of exponential asymptotic methods to resolve. Their theory relied upon studying the Stokes phenomenon exhibited by the capillary oscillations, which is induced by Gevrey divergence of the resurgent expansion that arises due to branch-point singularities in the analytic continuation of the leading-order gravity wave. 
The theory also produced a solvability condition, in which the $B \to 0$ asymptotic expansion broke down due to reordering at a countably infinite set of values of the Bond number that correspond to the locations between adjacent branches of solutions shown in figure~\ref{fig:gcsolspace}$(a)$.
These values of the Bond number are the fully nonlinear resonance condition first obtained in the linear regime by Wilton~\cite{wilton1915lxxii} and in the weakly-nonlinear regime by Chen \& Saffman~\cite{chen1979steady}.

\subsection{Subharmonic stability of solutions with $\mathcal{E}=0.0007$}
\label{sec:gcstab1}
\noindent We now calculate the linear stability of the steady solutions introduced in section~\ref{sec:gcsols}.
These solutions contained the effects of both gravity and surface tension, and were calculated at a fixed value of the energy \eqref{eq:energymap}.
We first discuss the stability of solutions with $\mathcal{E}=0.0007$.
Due to nondimensionalisation, the wavelength of the steady solution is equal to unity. The time-dependent perturbations are specified to have wavelength $1/p$, where $p$ is the Floquet exponent in solution \eqref{eq:floquetansatz}. 
For each value of $p$, there exist a set of eigenvalues $\sigma \in \mathbb{C}$ and eigenfunctions associated with the temporal stability of a specified steady solution.
The key quantities are the growth rate of the perturbation, Re$[\sigma]$, and the temporal frequency, Im$[\sigma]$.
In the following results, we calculate these quantities in the range $0 \leq p \leq 0.5$ using the numerical method detailed in Appendix~\ref{app:linear} to solve the generalised eigenvalue problem.

We first study the stability of the twenty solutions shown in figure~\ref{fig:gcsolspace}$(b)$.
Each of the solutions are located on consecutive branches in the $(B,F)$ parameter space with $\mathcal{E}=0.0007$.
We focus initially on the growth rate, Re$[\sigma]$, and the corresponding Floquet exponent $p$ for each positive growth rate.
This provides insight into the possible wavelengths ($1/p$) at which unstable modes exist.
\begin{figure}
\centering
\includegraphics[scale=1]{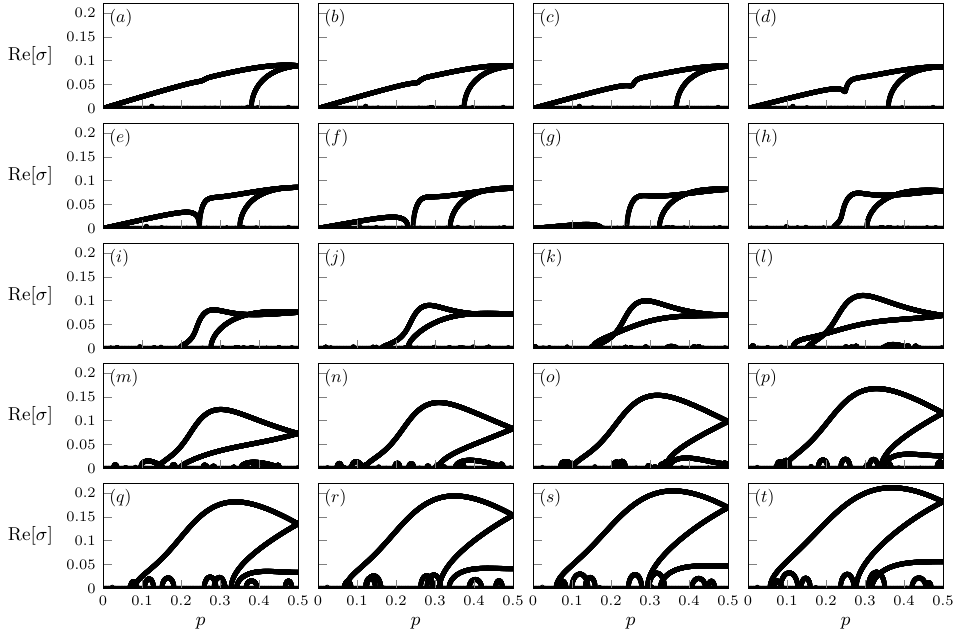}
\caption{\label{fig:gcsolspacesolsstab} Temporal stability results are shown for the twenty solutions from figure~\ref{fig:gcsolspace}$(b)$ with $\mathcal{E}=0.0007$, each of which is associated with a solution branch between $G_{28 \rightarrow 29}$ and $G_{9 \rightarrow 10}$.
The linear stability analysis calculates the time-dependent behaviour associated with different eigenfunctions, and the complex-valued eigenvalues $\sigma$ contribute to both growth (Re$[\sigma]$) and oscillations (Im$[\sigma]$) in time.
Panels $(a$--$t)$ show the growth rates, Re$[\sigma]$, and the respective Floquet exponent, $p$, associated with each mode. These results were calculated with the numerical scheme detailed in Appendix~\ref{app:linear}, and due to symmetry, only eigenvalues with Re$[\sigma] \geq 0$ are shown.}
\end{figure}
The stability of each of these solutions is shown in figure~\ref{fig:gcsolspacesolsstab}.
That shown in \ref{fig:gcsolspacesolsstab}$(a)$ is for $B = 0.0009537$ on branch $G_{9 \rightarrow 10}$, and that shown in \ref{fig:gcsolspacesolsstab}$(t)$ is for $B=0.002869$ on branch $G_{28 \rightarrow 29}$.
Both these values are reported to four significant digits.
All of these solutions are superharmonically stable as when $p=0$ all eigenvalues have Re$[\sigma]=0$.
Further, all solutions are subharmonically unstable as there exist values of $p$ at which the growth rate Re$[\sigma]$ is positive.
For solution $(a)$, unstable modes also exist at all values of $p>0$. 
This includes in the limit of $p \to 0$, and so this solution is modulationally unstable with respect to long-wave perturbations.
As the Bond number increases, the solutions become modulationally stable and this occurs with solutions $(h)$ to $(t)$.
While these solutions are modulationally stable to long wave perturbations, there remain values of $p$ at which subharmonic instability emerges.
For instance, solution $(h)$ is modulationally stable for $0<p \lessapprox 0.2$ and modulationally unstable for $0.2 \lessapprox p<0.5$.
As the surface tension increases, the associated subharmonic growth rates are also larger: solution $(m)$ has a maximum growth rate of Re$[\sigma]=0.1234$ at $p=0.302$ and solution $(t)$ has a maximum of Re$[\sigma]=0.2109$ at $p=0.369$.

The weakly-nonlinear theory described in section~\ref{sec:weaklynonlin} predicted only two transitions (away from resonant values of $B$) in the modulational stability at the values $B_1=0.003919$ and $B_2=0.01267$ defined in \eqref{eq:signvals}, such that the solutions are modulationally stable between $B_1<B<B_2$. 
The results in figure~\ref{fig:gcsolspacesolsstab} have demonstrated that the region of modulationally stability changes drastically as the amplitude of the solution increases from the weakly-nonlinear regime.
In particular, solutions $(h)$ to $(t)$ are stabilised at values of the Bond number $B$ well below the threshold predicted by weakly-nonlinear theory, and the transition to modulational instability between solutions $(h)$ and $(g)$ occurs below $B=0.001264$.

\begin{figure}
\centering
\includegraphics[scale=1]{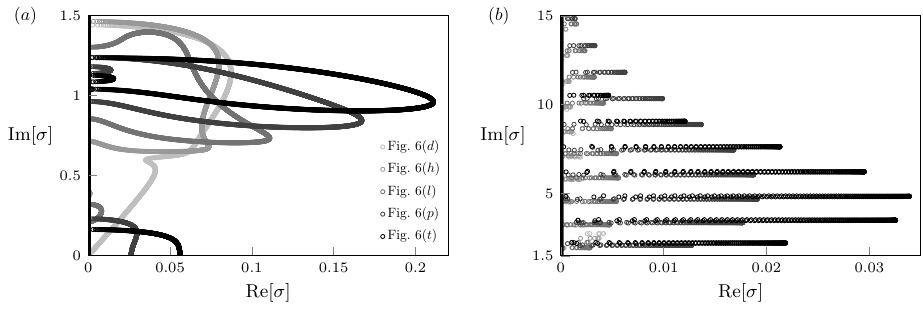}
\caption{\label{fig:gcsolspacesolsstabfull} The full eigenvalue spectrum is shown in the (Re$[\sigma]$,Im$[\sigma]$) plane for five gravity-capillary solutions studied in figure~\ref{fig:gcsolspacesolsstab} with $\mathcal{E}=0.0007$. Panel $(a)$ shows the behaviour of eigenvalues near the origin, and panel $(b)$ shows the behaviour at larger values of Im$[\sigma]$. The eigenvalues for each solution are shown with a different shade of grey as indicated in the legend. The location of these solutions in the $(B,F)$ parameter space was marked in figure~\ref{fig:gcsolspace}$(a)$, and have $B=\{0.001066,0.001265,0.001555,0.002017,0.002870\}$ to four significant figures.}
\end{figure}
The full stability spectrum is shown in the $(\text{Re}[\sigma],\text{Im}[\sigma])$ plane in figure~\ref{fig:gcsolspacesolsstabfull} for five of the stability results given previously in figure~\ref{fig:gcsolspacesolsstab}.
The region near the origin in shown in \ref{fig:gcsolspacesolsstabfull}$(a)$, and eigenvalues with larger imaginary part are shown in \ref{fig:gcsolspacesolsstabfull}$(b)$.
For the solution with smallest Bond number $B=0.001066$, shown in light grey, the eigenvalue spectrum passes through the origin which results in a typical figure-of-eight spectrum across all four quadrants much like that seen for the gravity wave with $\mathcal{E}=0.0007$ in figure~\ref{fig:gravstab}.
This figure-of-eight pattern splits up as the surface tension increases, which results in isolated bubbles along the imaginary axis.
These higher-frequency unstable modes become the dominant form of instability at larger values of the surface tension.
For instance, the solution with $B=0.002870$ shown in figure~\ref{fig:gcsolspacesolsstabfull}$(a)$ with black has the largest growth rate when Im$[\sigma]=0.961$. 
The eigenvalues further along the imaginary axis shown in \ref{fig:gcsolspacesolsstabfull}$(b)$ have a comparatively smaller growth rate.

\begin{figure}
\centering
\includegraphics[scale=1]{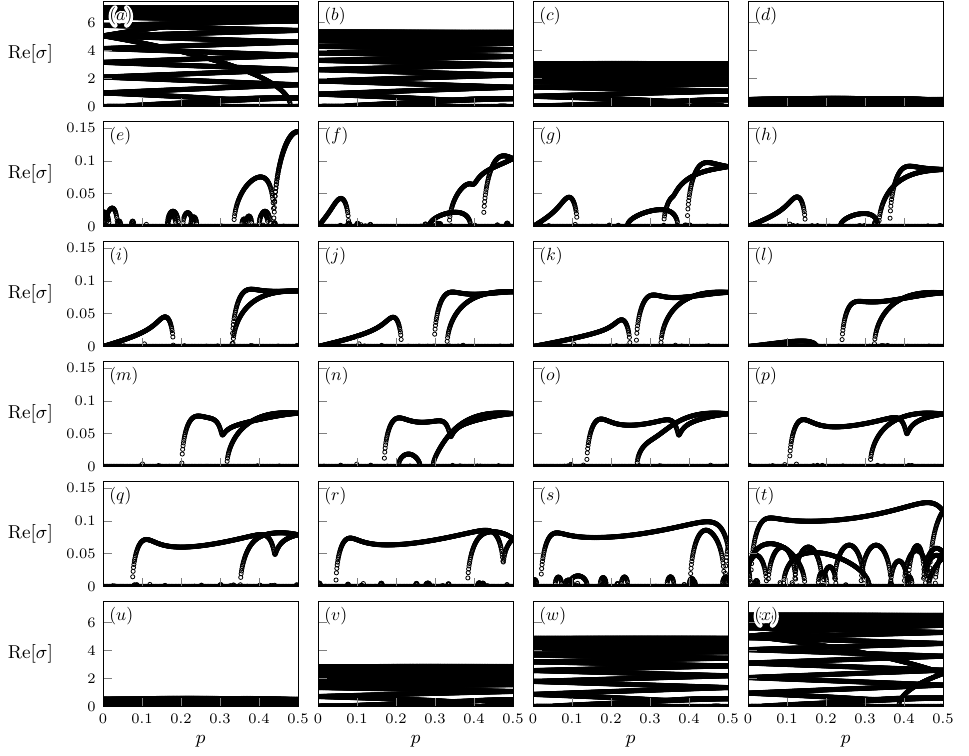}
\caption{\label{fig:finger1stab} Stability results for the twenty four marked solutions with $\mathcal{E}=0.0007$ from branch $G_{22 \rightarrow 23}$ in figure~\ref{fig:branches0007}$(a)$ are shown in the $(p,\text{Re}[\sigma])$ plane. The wavelength of the temporal perturbation is given by $1/p$. Solutions $(a$--$d)$ and $(u$--$x)$ correspond to the solutions on the sides of the bifurcation branch, and are highly unstable with respect to longitudinal perturbations of all possible wavelengths.}
\end{figure}
The stability of multiple solutions along the branch $G_{22 \rightarrow 23}$ for $\mathcal{E}=0.0007$ is shown in figure~\ref{fig:finger1stab}. 
The location of these twenty-four solutions in the $(B,F)$ plane was marked alongside the branch in figure~\ref{fig:branches0007}$(a)$.
We see that the four solutions along both ends of the branch, labelled $(a$--$d)$ and $(u$--$x)$, are highly unstable.
For these solutions, there exist eigenvalues with Re$[\sigma]>0$ for both $p=0$ and $0<p \leq 1/2$, and so these solutions are both superharmonically and subharmonically unstable.
The sixteen solutions along the top of the branch, labelled $(e$--$t)$ are superharmonically stable as Re$[\sigma]=0$ when $p=0$.
However, these solutions are all subharmonically unstable.
Solution \ref{fig:finger1stab}$(l)$ corresponds to the single solution on branch $G_{22 \rightarrow 23}$ previously studied in figure~\ref{fig:gcsolspacesolsstab}$(g)$, which was that with the largest surface tension to exhibit modulational instability.
We see from figure~\ref{fig:finger1stab} that the modulational stability properties of these solutions changes significantly along the individual branch.
Below a certain value of the surface tension the solutions \ref{fig:finger1stab}$(e$--$l)$ on the branch are modulationally unstable and above this value the solutions \ref{fig:finger1stab}$(m$--$t)$ are modulationally stable.
Note that while these latter solutions are stable in the long-wave limit, there can still exist perturbations of large wavelength that are unstable.
For instance, solution \ref{fig:finger1stab}$(q)$ is stable for $0 \leq p \lessapprox 0.075$ but unstable for $0.075 \lessapprox 0 \leq 0.5$ and so a perturbation with a wavelength ten times that of the travelling wave is unstable.

\begin{figure}
\centering
\includegraphics[scale=1]{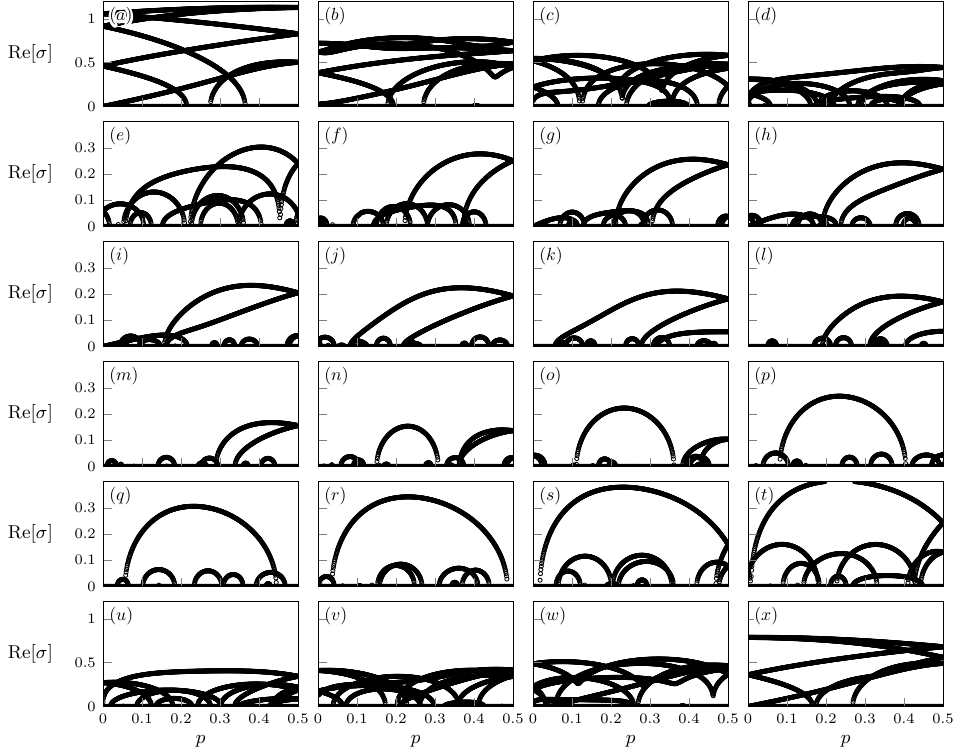}
\caption{\label{fig:finger2stab} Stability results for the twenty four marked solutions with $\mathcal{E}=0.0007$ from branch $G_{9 \rightarrow 10}$ in figure~\ref{fig:branches0007}$(c)$ are shown in the $(p,\text{Re}[\sigma])$ plane. The wavelength of the temporal perturbation is given by $1/p$. Solutions $(a$--$d)$ and $(u$--$x)$ correspond to the solutions on the sides of the bifurcation branch, and solutions $(e$--$t)$ to those along the top of the branch.}
\end{figure}
In figure~\ref{fig:finger2stab} stability results are shown for the solution branch $G_{9 \rightarrow 10}$ with $\mathcal{E}=0.0007$. These results are similar to that presented previously in figure~\ref{fig:finger1stab}, but for a branch that exists at larger values of surface tension.
Here, the solutions \ref{fig:finger2stab}$(a$--$f)$ and \ref{fig:finger2stab}$(t$--$x)$ are superharmonically unstable, and solutions \ref{fig:finger2stab}$(g$--$s)$ are superharmonically stable.
Note also that solutions $(g$--$i)$ are modulationally unstable, and so for a very small portion of the branch these is instability with respect to long-wave perturbations.
Further, in many of the stability results there are small regions of $p$ at which unstable modes exist but with smaller values of Re$[\sigma]$; these correspond to higher-frequency instabilities such as those shown in figure~\ref{fig:gcsolspacesolsstabfull}$(b)$.
\subsection{Superharmonic and modulational instabilities}
\label{sec:supermod}
\begin{figure}
\centering
\includegraphics[scale=1]{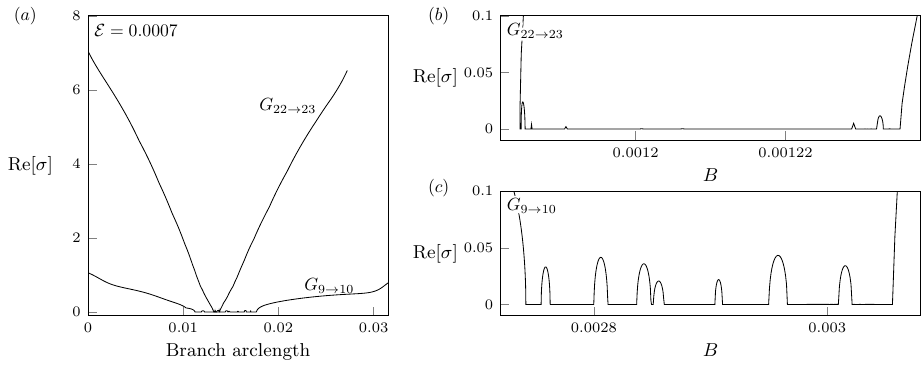}
\caption{\label{fig:super0007} Stability results are shown for superharmonic perturbations, which have $p=0$ in the eigenfunction \eqref{eq:floquetansatz}, to solutions with $\mathcal{E}=0.0007$. Solutions are studied along the two branches $G_{22 \rightarrow 23}$ and $G_{9 \rightarrow 10}$ shown previously in figure~\ref{fig:branches0007} and studied in figures~\ref{fig:finger1stab} and \ref{fig:finger2stab}. Panel $(a)$ shows the largest growth rate with $p=0$ for each solution along the branch, parameterised by the branch arclength. Solutions are superharmonically unstable at the start and end of each branch, with regions in the middle that are superharmonically stable. The stable regions along each branch are displayed in panels $(b)$ and $(c)$, which shows the maximum growth rate against the corresponding Bond number, $B$, for each solution.}
\end{figure}
\noindent We demonstrated in section~\ref{sec:gcstab1} that along each solution branch there is a change (and consequently stabilisation) in both the superharmonic and modulational instabilities.
It is now investigated how these transitions occur along each solution branch.
Firstly, in figure~\ref{fig:super0007} we study the superharmonic instability with $p=0$.
The largest growth rate is shown in \ref{fig:super0007}$(a)$ along the two branches $G_{22 \rightarrow 23}$ and $G_{9 \rightarrow 10}$, measured by the branch arclength.
Along the start and end of each branch there is superharmonic instability, which confirms our previous assertions made when analysing the stability spectrum along each branch in figures~\ref{fig:finger1stab} and \ref{fig:finger2stab}.
The largest superharmonic growth rate is shown along the top of each branch in figure~\ref{fig:super0007}$(b,c)$.
Here, the solutions are mostly superharmonically stable.
However, we observe that there exist small regions of instability that are more prominent for the branch $G_{9 \rightarrow 10}$.
These unstable superharmonic modes have high frequency, Im$[\sigma]$.

\begin{figure}
\centering
\includegraphics[scale=1]{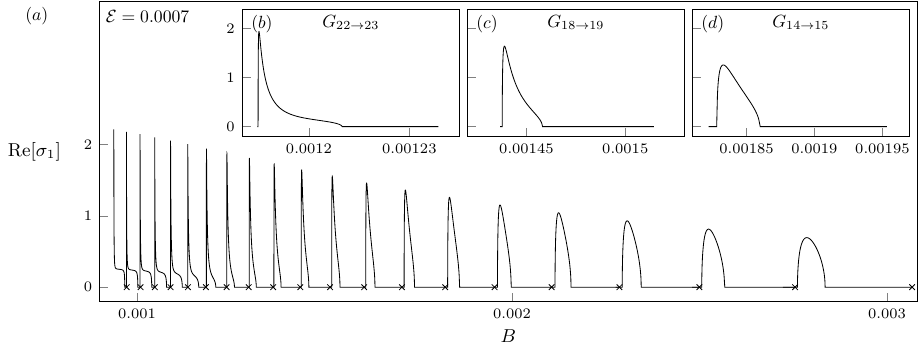}
\caption{\label{fig:mod0007} Stability results are shown for long-wave (modulational) perturbations in the limit of $p \to 0$ to solutions with energy $\mathcal{E}=0.0007$. The solutions studied are those at the top of each solution branch in figure~\ref{fig:gcsolspace} that are predominantly superharmonically stable, such as the two ranges of $B$ shown in figure~\ref{fig:super0007}$(b,c)$. Specifically, we plot the real part of $\sigma_1=\sigma/p$ calculated with $p=0.0005$. This corresponds to a numerical estimate of the long-wave expansion of the eigenvalue $\sigma \sim \sigma_0 + \sigma_1 p + O(p^2)$, when the solution is superharmonically stable with Re$[\sigma_0]=0$. The insets $(b$--$d)$ show an enhanced view of the long-wave growth rate across the three solution branches $G_{22 \rightarrow 23}$, $G_{18 \rightarrow 19}$, and $G_{14 \rightarrow 15}$.}
\end{figure}
The behaviour of the long-wave modulational instability is shown in figure~\ref{fig:mod0007}$(a)$ for branches across the range $0.001<B<0.003$.
Here, we numerically estimated the long-wave growth rate in the limit of $p \to 0$, $\sigma \sim 0 + \sigma_1 p + O(p^2)$, by the formula $\sigma_1=\sigma/p$ evaluated with $p=0.0005$.
We see that along each branch studied there is a transition between modulational stability and instability.
This is shown more clearly in each of the three insets.
For the branch $G_{22 \rightarrow 23}$ shown in figure~\ref{fig:mod0007}$(b)$, the transition in stability is approximately half way along the top of the branch. Solutions are then modulationally unstable on the left-hand side within $0.00118<B<0.00121$, and stable on the right-hand side within $0.00121<B<0.00124$.
For branches obtained at higher values of surface tension, such as those shown in \ref{fig:mod0007}$(c,d)$, fewer solutions along the branch are modulationally unstable. 
Then, as the Bond number deceases (towards smaller surface tension), more solutions along the branch become modulationally unstable.
We have therefore shown that the effect of nonlinearity is both to stabilise solutions with respect to long-wave perturbations well below the threshold value predicted by weakly-nonlinear theory in section~\ref{sec:weaklynonlin}, and that this stabilisation is nonmonotonic.

\subsection{Subharmonic stability of solutions with $\mathcal{E}=0.0015$}
\label{sec:gcstab2}
\noindent We now consider the stability of gravity-capillary waves obtained with a higher energy value, $\mathcal{E}=0.0015$.
In the absence of surface tension, the $\mathcal{E}=0.0015$ Stokes wave is both superharmonically and modulationally stable, but is subject to subharmonic instability with a maximum growth rate of Re$[\sigma]=0.1288$ at $p=0.36$ as shown in figure~\ref{fig:gravstab}$(c)$.

\begin{figure}
\centering
\includegraphics[scale=1]{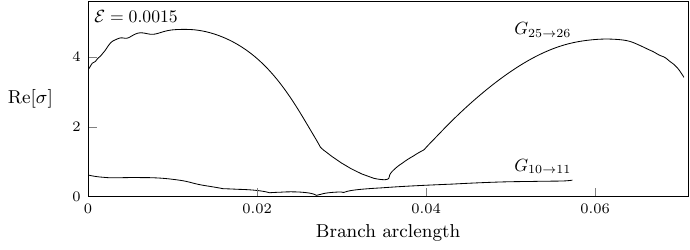}
\caption{\label{fig:super0015} Superharmonic stability results, which have $p=0$ in the eigenfunction \eqref{eq:floquetansatz}, are shown for solutions with $\mathcal{E}=0.0015$. Solutions are studied along the two branches $G_{25 \rightarrow 26}$ and $G_{10 \rightarrow 11}$ shown previously in figure~\ref{fig:branches0015}. The largest growth rate (with $p=0$) is shown for each solution along the respective branch, parameterised by the branch arclength. Solutions are superharmonically unstable throughout the entire branch.}
\end{figure}
First, we consider the stability of superharmonic perturbations with $p=0$. Superharmonic growth rates are shown in figure~\ref{fig:super0015} for the two branches $G_{25 \rightarrow 26}$ and $G_{10 \rightarrow 11}$.
These branches and corresponding solution profiles were shown
figure~\ref{fig:branches0015}, and correspond to solutions near $B=0.0013$ for branch $G_{25 \rightarrow 26}$ and near $B=0.0028$ for $G_{10 \rightarrow 11}$.
All solutions along these branches are found to be superharmonically unstable, with the largest growth rate being minimal at the top of each solution branch (where the wavespeed is highest).

\begin{figure}
\centering
\includegraphics[scale=1]{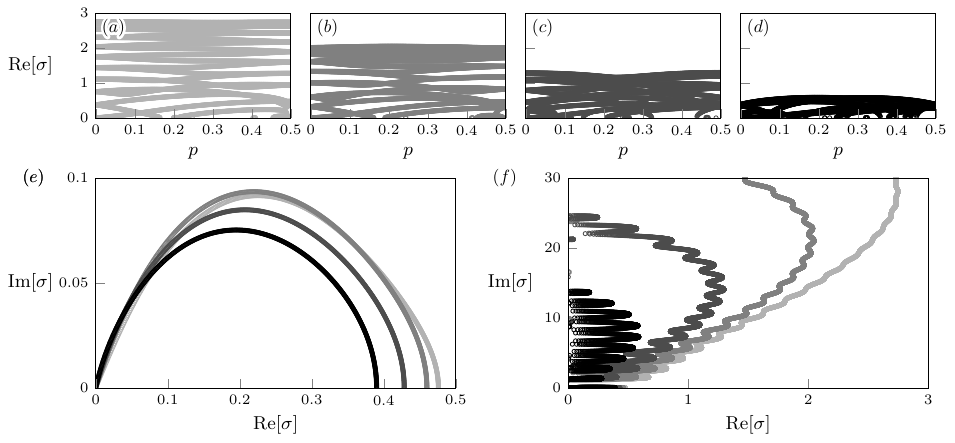}
\caption{\label{fig:stab0015} Stability results are shown for four solutions with $\mathcal{E}=0.0015$.
These solutions, which have $B=\{0.001068,0.001350,0.001833,0.002872\}$ and lie along branches $G_{28 \rightarrow 29}$, $G_{22 \rightarrow 23}$, $G_{16 \rightarrow 17}$, and $G_{10 \rightarrow 11}$, respectively, are the solution with the smallest capillary ripple amplitude along the branch.
Panels $(a$--$d)$ show the dependency on the growth rates, Re$[\sigma]$ against the Floquet exponent $p$. The full eigenvalue spectrum, $(\text{Re}[\sigma],\text{Im}[\sigma])$, is shown near the origin in panel $(e)$ and for larger values of Im$[\sigma]$ in panel $(f)$.}
\end{figure}
The full eigenvalue spectrum for subharmonic perturbations to four individual solutions with $\mathcal{E}=0.0015$ is shown in figure~\ref{fig:stab0015}.
These solutions are taken from the branches $G_{28 \rightarrow 29}$, $G_{22 \rightarrow 23}$, $G_{16 \rightarrow 17}$, and $G_{10 \rightarrow 11}$, and were selected as being the solution with smallest capillary ripple amplitude along each branch.
Each of these solutions are both superharmonically and subharmonically unstable as seen in \ref{fig:stab0015}$(a$--$d)$.
Further, it is solution \ref{fig:stab0015}$(a)$ which has the smallest value of surface tension studied that is the most unstable here.
The eigenvalues close to the origin are shown in figure~ \ref{fig:stab0015}$(e)$, and those with higher frequencies shown in \ref{fig:stab0015}$(f)$.
While there are unstable modes close to the origin with $0<\text{Im}[\sigma]<0.1$ in \ref{fig:stab0015}$(e)$ for each solution, it is the high-frequency instabilities shown in \ref{fig:stab0015}$(f)$ that have the largest growth rates.
The largest growth rate for solution \ref{fig:stab0015}$(a)$ occurs for $p=0.045$ and has both Re$[\sigma]=2.748$ and the very high temporal frequency Im$[\sigma]=28.02$.
However there are unstable modes with Re$[\sigma]>2.729$ for all values of $0 \leq p \leq 1/2$, and so the wavelength of the subharmonic perturbation is less significant for these solutions.
These results suggest that the high-frequency instability is the dominant instability for very high-amplitude gravity-capillary waves.

\section{Conclusion and discussion}
\label{sec:disc}
\noindent We have studied the linear temporal stability of travelling gravity-capillary waves at large values of the solution amplitude.
One of the main difficulties in our analysis is the complexity of the underlying solution space that emerges for non-zero surface tension, which takes the form of a countably infinite set of branches that emerge in the small-surface-tension limit.
Our results have included stability with respect to superharmonic and subharmonic perturbations, both captured by specifying the perturbation wavelength via the Floquet exponent.
In the absence of surface tension, gravity waves are subharmonically unstable (which in the long-wave limit is known as the Benjamin-Feir or modulational instability) and exhibit superharmonic instability only close to the Stokes wave of limiting form.

We have demonstrated that even a small amount of interfacial tension significantly alters the stability properties of travelling surface waves.
We first studied the stability of nonlinear solutions defined with energy $\mathcal{E}=0.0007$, which corresponds to the energy of a gravity wave of approximately half the limiting amplitude.
The gravity wave with $\mathcal{E}=0.0007$ exhibits the well-known modulational instability, and we have discovered that this can be suppressed with the inclusion of surface tension.
Firstly, in the weakly-nonlinear regime we derived the critical ranges of the nondimensional Bond number $0.003919 < B < 0.01267$ within which the long-wave instability is stabilised.
Then, in the fully nonlinear regime at $\mathcal{E}=0.0007$ we demonstrated that this threshold changes even further with modulational stable solutions existing above the value of $B=0.001$.
However, the modulational stability of these is nonmonotonic and changes across each individual solution branch.
Therefore, not only does the presence of surface tension influence the long-wave stability, but even very small changes in the surface tension were seen to alter the stability properties of solutions.
We also studied solutions with $\mathcal{E}=0.0015$, corresponding to the energy of a gravity wave with approximately eighty percent the of limiting amplitude.
All gravity-capillary waves studied at this energy level were superharmonically unstable and subject to high-frequency instabilities.

In our results we focused on the interval $0.001<B<0.003$.
By using the definition of the Bond number $B$ defined in \eqref{eq:nondimparams}, we can use typical properties of water and gravity on the earths surface to determine the corresponding wavelengths for which our results are valid. 
From $\sigma=0.07197 \cdot \text{Kg} \cdot \text{s}^{-2}$ for the coefficient of surface tension, $\rho=1000 \cdot \text{Kg}\cdot \text{m}^{-3}$ for the fluid density, and $g=9.81 \cdot \text{m} \cdot \text{s}^{-2}$ for gravitational acceleration, we find that the interval $0.001<B<0.003$ corresponds to wavelengths $0.0495 \cdot \text{m}<\lambda <  0.0857\cdot\text{m}$.
To obtain results valid for waves of larger wavelengths one would have to study solutions with smaller values of the Bond number.
We note that the stability of finite-amplitude capillary waves has been studied by Blyth \& P{\u{a}}r{\u{a}}u \citep{blyth2016stability}, however this is a distinct regime to that we have focused on as pure capillary waves emerge in our formulation in the limit of $F \to \infty$ with $B=O(F^2)$.

In the absence of surface tension, gravity waves are known to be superharmonically unstable at very high amplitudes close to the Stokes wave of limiting form \cite{tanaka1983stability}.
The instability near the limiting Stokes wave was recently studied in detail by Deconinck~\emph{et al.}~\cite{deconinck2024self}.
They observed numerically that under the approach to limiting steepness unstable eigenvalues are generated at the origin associated with modulational instability which then detach along the axis of the complex plane, and that this process repeats indefinitely and is associated with a countably infinite number of turning points of the wavespeed and energy with respect to amplitude. 
This behaviour was subsequently studied rigorously by Dyachenko~\emph{et al.}~\cite{dyachenko2026recurrent}.

High-frequency modes, analogous to those that we observed to be the dominant form of instability for gravity-capillary waves with $\mathcal{E}=0.0015$ in section~\ref{sec:gcstab2}, have been well studied for gravity waves.
It was shown by Deckoninck \& Oliveras~\cite{deconinck2011instability} that for gravity waves the high-frequency modes have a higher growth rate than the modulational modes at certain ranges of the fluid depth. 
These high-frequency modes have been derived analytically by Creedon \emph{et al.}~\cite{creedon2022high} in the small-amplitude limit.

Our results have focused only on the stability with respect to longitudinal perturbations whose shape changes only in the direction of wave propagation.
Travelling surface waves are known to be unstable with respect to general three-dimensional perturbations, with independent wavenumbers existing in both the longitudinal and transverse directions. 
The transverse instability of gravity waves was first shown numerically by McLean~\cite{mclean1982instabilities}, and the resultant unstable modes are frequently observed in experiments \cite{trulsen2025experimental}.
The transverse instability of gravity-capillary waves on finite depth has been studied by \cite{haragus2023transverse}, but for small amplitude travelling waves and perturbations that are superharmonic in the longitudinal direction.

The results of linear stability theory that we have so far discussed are all formally valid in the limit of small time.
In time-evolution problems, the effect of nonlinearity and even nonnormality in the linear stability problem can significantly alter the observed behaviour.
Starting from an initial condition of a nonlinear gravity wave, the effect of surface tension has been studied in superharmonic simulations by both Hung \& Tsai~\cite{hung2009formation} and Murashige \& Choi \cite{murashige2017numerical}, who observed that oscillatory capillary modes initially form in an asymmetric manner ahead of the wave crest before spreading to the remainder of the wave profile and settling into general unsteady motion.
These capillary modes often have a high temporal frequency, as observed by Shelton \emph{et al.}~\cite{shelton2023structure} for time-periodic gravity-capillary standing waves.
Further, superharmonic modes were studied in a gravity-capillary formulation including model viscosity and wind forcing by Shelton \emph{et al.}~\cite{shelton2025time} where nonlinear stability was demonstrated for the initial conditions used.

\begin{acknowledgments}
\noindent Part of this research was conducted during a visit to the Department of Mathematics at The Pennsylvania State University in 2025, and JS would like to thank Prof. Paul Milewski for hospitality and valuable discussions during this visit.
\end{acknowledgments}

\appendix
\section{Details of the numerical implementations}
\noindent We now detail the numerical implementation of both the steady problem and the linear stability problem. Matlab code that implements each of these is provided in our supplementary material.
\subsection{Steady solutions}
\label{app:steady}
\noindent Solutions of the gravity-capillary wave problem that are steady in a comoving frame satisfy equation \eqref{eq:steadybern}. To obtain these numerically, we consider the Fourier series decomposition
\begin{equation}\label{eq:fourierdecomp}
Y(\xi)= \sum_{n=0}^{N}a_n \cos{(2 n \pi \xi)},
\end{equation}
which yields $N+2$ unknown constants to be determined: $N+1$ Fourier coefficients from \eqref{eq:fourierdecomp} and the Froude number, $F$.
This is closed by $N+2$ equations. One of these is the energy constraint \eqref{eq:energymap}, and the remaining $N+1$ equations come from evaluating equation \eqref{eq:steadybern} in Fourier space.
In order to evaluate these equations, we first specify values for $B$ and $\mathcal{E}$, and make a guess for the unknowns $\{ a_0,\ldots,a_N,F\}$.
This guess is either a previously calculated numerical solution (obtained with different parameter values), or a small-amplitude analytical solution given by $\{0, \epsilon ,0 \ldots ,0, \sqrt{(1+4 \pi^2 B)/(2 \pi)}\}$, where $\epsilon$ is small.

To evaluate each algebraic equation, we calculate $X(\xi)$ from harmonic relation \eqref{eq:harmonic1} by using the Fourier multiplier of the Hilbert transform, obtained from $\mathcal{H}[\mathrm{e}^{2 n \pi \mathrm{i} \xi}]=\mathrm{i} \cdot \text{sgn}(n)\mathrm{e}^{2 n \pi \mathrm{i} \xi}$. Derivatives are also calculated spectrally by noting that the Fourier multiplier for differentiation is $2 n \pi \mathrm{i}$.
Newton iteration is then used to minimise the residual of this system, measured by the $L^2$ norm, to below a tolerance of $10^{-11}$. 
This method typically converges to a new numerical solution provided that the initial guess is reasonably close to the true solution at the specified parameter values. 
For instance, in the continuation program used to locate branches of numerical solutions in the $(B,F)$ bifurcation space, very small jumps are taken along the arclength of each branch.

\begin{figure}
\centering
\includegraphics[scale=1]{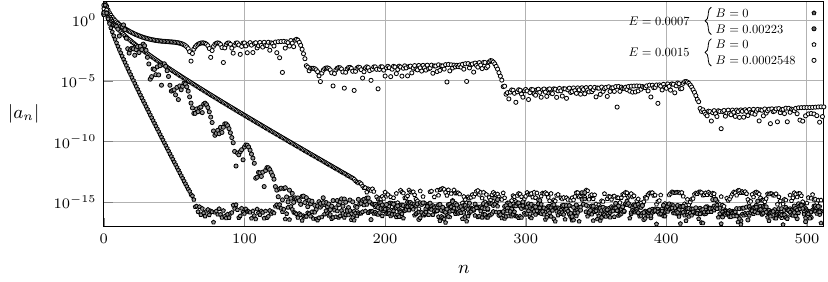}
\caption{\label{fig:fourier} The Fourier spectrum of four numerical solutions to equation \eqref{eq:steadybern} are shown. The two solutions with $E=0.0007$ and that with $E=0.0015$ and $B=0$ were computed with $N=511$ Fourier coefficients in decomposition \eqref{eq:fourierdecomp}. The solution with $E=0.0015$ and $B=0.0002548$ was computed with $N=1024$.}
\end{figure}

For a solution with $N=511$, convergence takes approximately one second on a desktop computer. However, obtaining the growth rates for the linear stability of a solution with $N=511$ modes for a range of $501$ values of the Floquet exponent $p$ (as detailed in section~\ref{app:linear}) takes approximately 8 hours.
Thus, to save computational time we further truncate the number of Fourier modes in the steady solution when calculating the linear stability growth rates.
Specifically, if there exists a value of $n$ beyond which all Fourier coefficients lie below $10^{-13}$, then this forms the truncation point.
The Fourier coefficients, $a_n$, of four numerical solutions with $N=512$ are shown in figure~\ref{fig:fourier} as an example of this.
The solution with $E=0.0007$ and $B=0$ would be truncated to $N=55$, that with $E=0.0007$ and $B=0.00223$ to $N=121$, and that with $E=0.0015$ and $B=0$ to $N=177$. 
The solution with $E=0.0015$ and $B=0.0002548$ would be recomputed with $N=2047$.

\subsection{Linear stability}
\label{app:linear}
\noindent Given a numerical solution of the steady problem obtained with $N$ Fourier coefficients via decomposition \eqref{eq:fourierdecomp}, the Fourier series expansion for the perturbation is truncated similarly via
\begin{equation}\label{eq:floquetansatztruncate}
\{ X_1,Y_1,\Phi_1,\Psi_1 \}=\mathrm{e}^{\sigma t} \sum_{n=-N}^{N} \{a_n,b_n,c_n,d_n \} \mathrm{e}^{2 (n+p) \pi \mathrm{i} \xi}.
\end{equation}
Upon using the identity $\mathcal{H}[\mathrm{e}^{2 (n+p) \pi \mathrm{i} \xi}]=\mathrm{i} \cdot \text{sgn}(n+p)\mathrm{e}^{2 (n+p) \pi \mathrm{i} \xi}$, the $O(\epsilon)$ harmonic relations (\hyperref[eq:harmonic1Oep]{\ref{eq:harmonic1Oep},d}) immediately give $b_n=- \mathrm{i} \cdot \text{sgn}(n+p)a_n$ and $d_n=\mathrm{i} \cdot \text{sgn}(n+p)c_n$, such that only the constants $a_n$ and $c_n$ need to be considered as unknowns of the problem. Thus, we consider the solution vector
\begin{equation}
\mathbf{v}=(a_{-N},\ldots,a_{N},c_{-N},\ldots,c_{N})^{T},
\end{equation}
which contains $4N+2$ entries, and where ${}^T$ denotes transpose. 
We introduce the $2N+1$ collocation points defined by $\xi_i=(i-1)/(2N+1)-1/2$ for $i=1,\ldots,2N+1$.
Then, by evaluating the two governing equations (\hyperref[eq:YevolveOep]{\ref{eq:YevolveOep},b}) at each of these collocation points, we obtain $4N+2$ algebraic equations in total which forms a closed system for the algebraic unknowns.
By collecting these algebraic equations into a matrix system, we obtain the generalised eigenvalue problem
\begin{equation}\label{eq:eigenvalue}
\sigma \mathbf{L} \mathbf{v}= \mathbf{R} \mathbf{v},
\end{equation}
where $\mathbf{L}$ and $\mathbf{R}$ are $(4N+2) \times (4N+2)$ matrices.
The entries of $\mathbf{L}$ arise from the time-derivative terms $Y_{t}$ and $\Phi_{t}$, and are given by $\mathbf{L}_{ij}=\mathrm{e}^{2 (p+j-N-1) \pi \mathrm{i} \xi_i}$ for $1 \leq i \leq 2 N+1$ and $1 \leq j \leq 2 N+1$,
 $\mathbf{L}_{ij}=\mathrm{e}^{2 (p+j-3N-2) \pi \mathrm{i} \xi_{i-N}}$ for $2N+2 \leq i \leq 4N+2$ and $2N+2 \leq j \leq 4N+2$, and $\mathbf{L}_{ij}=0$ otherwise.
The construction of the matrix $\mathbf{R}$ is detailed in the code in our supplementary material, and involves terms arising from the right-hand sides of equations (\hyperref[eq:YevolveOep]{\ref{eq:YevolveOep},b}), for which further details are given by \cite{blyth2016stability,tiron2012linear}.

The generalised eigenvalue problem \eqref{eq:eigenvalue} is solved in Matlab with the inbuilt function eig, which uses the QZ algorithm. For a solution with $N=512$ and a fixed value of $p$, this takes approximately 60 seconds to run on a desktop computer. Our results in section \ref{sec:gcstab} repeated this process for $501$ equally spaced values of $p$ in the range $p \in [0,0.5]$, which takes approximately $8$ hours to run in total. 
As detailed at the end of section~\ref{app:steady}, this runtime can be reduced provided that higher Fourier coefficients of the steady solution are at computer precision, in which case the total number of Fourier coefficients in the steady solution may be truncated further.
Since the runtime of the QZ algorithm is of $O(N^3)$, reducing the value of $N$ yields considerable improvements in run time.

\bibliography{GCbib}
\end{document}